\pdfoutput=1

\documentclass[fleqn,usenatbib]{mnras}

\usepackage{newtxtext,newtxmath}

\usepackage[T1]{fontenc}
\usepackage{ae,aecompl}
\usepackage{threeparttable}
\usepackage{blindtext}
\usepackage{natbib}

\usepackage{graphicx}	
\usepackage{amsmath}	
\usepackage{amssymb}	

\title[Vertical Na-O in NGC~6553]{The vertical Na-O relation in the  bulge globular cluster NGC~6553}

\author[C. Mu\~noz et al.]{
C. Mu\~noz$^{}$\thanks{E-mail:cesar.alejandro.munoz.g@gmail.com},
S. Villanova$^{1}$
D. Geisler$^{1,4,5}$,
C.C. Cort\'es$^{1,3}$,
C. Moni Bidin$^{2}$,
R.E. Cohen $^{7}$,
\newauthor
I. Saviane $^{6}$,
B. Dias$^{6,9}$,
B. Tang$^{8}$,
F. Mauro$^{2}$
\\
\\
$^{1}$Departamento de Astronom\'ia, Casilla 160-C, Universidad de
  Concepci\'on, Concepci\'on, Chile.\\
$^{2}$Instituto de Astronom\'ia, Universidad Cat\'olica del Norte, Av. Angamos 0610, Antofagasta, Chile.\\
 $^{3}$Departamento de F\'isica, Facultad de Ciencias, Universidad del B\'io-B\'io, Avenida Collao 1202, Casilla 15-C, Concepci\'on, Chile.\\
$^{4}$Instituto de Investigaci\'on Multidisciplinario en Ciencia y Tecnolog\'ia, 
Universidad de La Serena. Avenida Ra\'ul Bitr\'an S/N, La Serena, Chile.\\
$^{5}$Departamento de  Astronom\'ia, Facultad de Ciencias, Universidad de La Serena. Av. Juan Cisternas 1200, La Serena, Chile.\\
$^{6}$ European Southern Observatory, Casilla 19001, Santiago, Chile.\\
$^{7}$ Space Telescope Science Institute, 3700 San Martin Drive, Baltimore, MD 21218, USA.\\
$^{8}$ School of Physics and Astronomy, Sun Yat-sen University, Zhuhai 519082, People's Republic of China.\\
$^{9}$Departamento de F\'isica, Facultad de Ciencias Exactas, Universidad Andr\'es Bello, Av. Fernandez Concha 700, Las Condes, Santiago, Chile.
}

\date{Accepted XXX. Received YYY; in original form ZZZ}

\pubyear{2015}

\begin{document}
\label{firstpage}
\pagerange{\pageref{firstpage}--\pageref{lastpage}}
\maketitle

\begin{abstract}
In this article, we present a detailed chemical analysis of seven red giant members of NGC~6553 using high-resolution spectroscopy from VLT FLAMES. We  obtained the stellar parameters ($T_{eff}$, Log(g), $v_{t}$, [Fe/H]) of these stars from the spectra, and we measured the chemical abundance for 20 elements, including light elements, iron-peak elements, $\alpha$-elements and neutron-capture elements. The metallicities in our sample stars are consistent with a homogeneous distribution. We found a  mean of [Fe/H]=-0.14$\pm$0.07 dex, in agreement with other studies. Using the alpha-elements Mg, Si, Ca and Ti we obtain the mean of  [$\alpha$/Fe]=$0.11\pm0.05$. We found a vertical relation between Na and O, characterized by a significant spread  in Na and an almost non-existent spread in O. In fact, Na and Al are the only two light elements with a large intrinsic spread, which demonstrates the presence of Multiple Populations (MPs). An intrinsic spread in Mg is not detected in this study.  The $\alpha$, iron-peak and neutron capture elements show good agreement with the trend of the bulge field stars, indicating similar origin and evolution, in concordance  with our previous studies for two other  bulge GCs (NGC~6440 and NGC~6528).

\end{abstract}




\section{Introduction}
Having a complete picture of  different components of our Galaxy will allow us to understand with more detail their  formation, evolution  and the different astrophysical processes which have been involved during their lifetime. In fact,  nowadays,  our knowledge regarding  the Milky  Way has  been  greatly  improved, thanks to a large number researchers and new surveys which use the latest generation of facilities such as the VLT FLAMES,  VVV/VVVX survey \citep {minniti10}, the Gaia-ESO survey \citep {gilmore12}, SDSS-IV \citep {blanton17}, and the Gaia mission.

In this picture, undoubtedly the bulge of the Milky Way occupies an essential place. For this reason, there are more and more researchers studying distinct components or astrophysical  processes in the bulge, such as dynamics \citep{Portail2017, Beaulieu2000}, chemistry \citep{Nandakumar2018, Grieco2012}, and  even exoplanets \citep{ccortes18, Sahu2006}, among others. 
Since the bulge is likely the oldest component of our Galaxy, it can give us relevant information on its formation and subsequent evolution. One of the fundamental constituents of the bulge are  globular clusters (GCs), less studied than their  halo counterparts  due to difficulties including high and often variable extinction, even  across the small angular extent of a typical GC, as well as crowding and the difficulty in separating true bulge stars from intervening thin and thick disk stars.

Many studies have analyzed in detail the GCs of our Galaxy,  using high, medium and low-resolution spectroscopy, and photometry in many different bands. However  most of these studies mainly focus on the more accessible sections of our Galaxy, avoiding in large part the bulge. These studies have uncovered an intrinsic intracluster variation in a variety of light elements, including C, N, O, Na, Mg, Al and Si, which has become the major manifestation of the phenomenon known as Multiple Populations in GCs. MPs have been found in all galactic GCs, with only a few exceptions such as Ruprecht~106 \citep{villanova13,dotter18}. Moreover, most of these studies have focused on the detection of Na-O and Mg-Al anticorrelation \citep{carretta09b},  which are related to several nucleosynthesis processes whose ejecta might then allow the formation of a new generation of stars in the GCs with a consequent spread in the relevant light element abundances.

Another exciting aspect is the spread in iron found in some galactic  GCs    \citep {johnson08, marino11a, marino11b, dacosta09, carretta10a, origlia11}. However, this is still an open question that needs to be clarified, since in some GCs this spread is not clear, e.g., in the case of NGC~3201 \citep{munoz13,mucciarelli15}, or in other cases where  iron spread  has been found with  more uncertain techniques. Indeed, \citet{mauro14} found indications of significant spreads in some bulge GCs using Calcium Triplet technique in combination with NIR photometry. We have investigated these claims in two previous studies \citep{munoz17,munoz18}, and here investigate a third cluster in this respect, NGC~6553.

In our initial papers on bulge GCs, we performed chemical tagging of NGC~6440 \citep{munoz17} and NGC~6528 \citep{munoz18}. We found a very short extension in the Na-O anticorrelation, more consistent with a significant spread only in Na but no in O, and a chemical evolution somewhat different from their halo counterparts, but in  agreement with the chemical evolution of the bulge of our Galaxy.

In an effort by our group to expand the study of the bulge GCs, we present in this article a detailed chemical analysis for the bulge GC NGC~6553. This is clearly a bulge GC, located at a distance of only  2.2 Kpc\citep[2010 edition]{harris96} from the Galactic center. Like most  bulge GCs, it has  a high nominal reddening of  E(B-V)= 0.63 \citep[2010 edition]{harris96}, with a complex differential reddening.

 NGC~6553 has been  the subject of several studies using a variety of techniques in different wavelengths. For example,  \citet{dias16a} studied this GC  among other galactic GCs, using low resolution spectroscopy. \citet{tang17} studied NGC~6553 using high-resolution spectroscopy  from APOGEE \citep{majewski17}  part of the Sloan Digital Sky Survey III \citep{eisenstein11}. Also, \citet{cohen17} and \citet{mauro14} performed  Point Spread Function (PSF)  photometry  for a set of bulge GCs, which include NGC~6553, using data  from  the VVV survey \citep{minniti10}. Indeed, in this article, we take advantage of this photometry for use in target selection.

In the next section, we describe our observations and data reduction procedure, in section three we describe in detail the method used to obtain the atmospheric parameters, errors and chemical abundances. In section four we present our results regarding  iron-peak elements, alpha-elements, Na-O anti-correlation, Mg-Al-Na relations, and neutron capture elements. Finally, in section five, we summarize  our main findings.
\section{Observations and data reduction}
 
We observed seven red giants stars in NGC~6553  with the fiber-fed multiobject FLAMES spectrograph  mounted at the ESO VLT/UT2 telescope in Cerro Paranal (Chile) in period 93A (ESO program ID 093.D-0286, PI S. Villanova).  The analyses of the stars observed with FLAMES were conducted using the blue and red arms of the high-resolution spectrograph UVES. We obtained a single spectra for each star with a exposure time of 2774 seconds.

\begin{figure}
  \includegraphics[width=3.4in,height=3.4in]{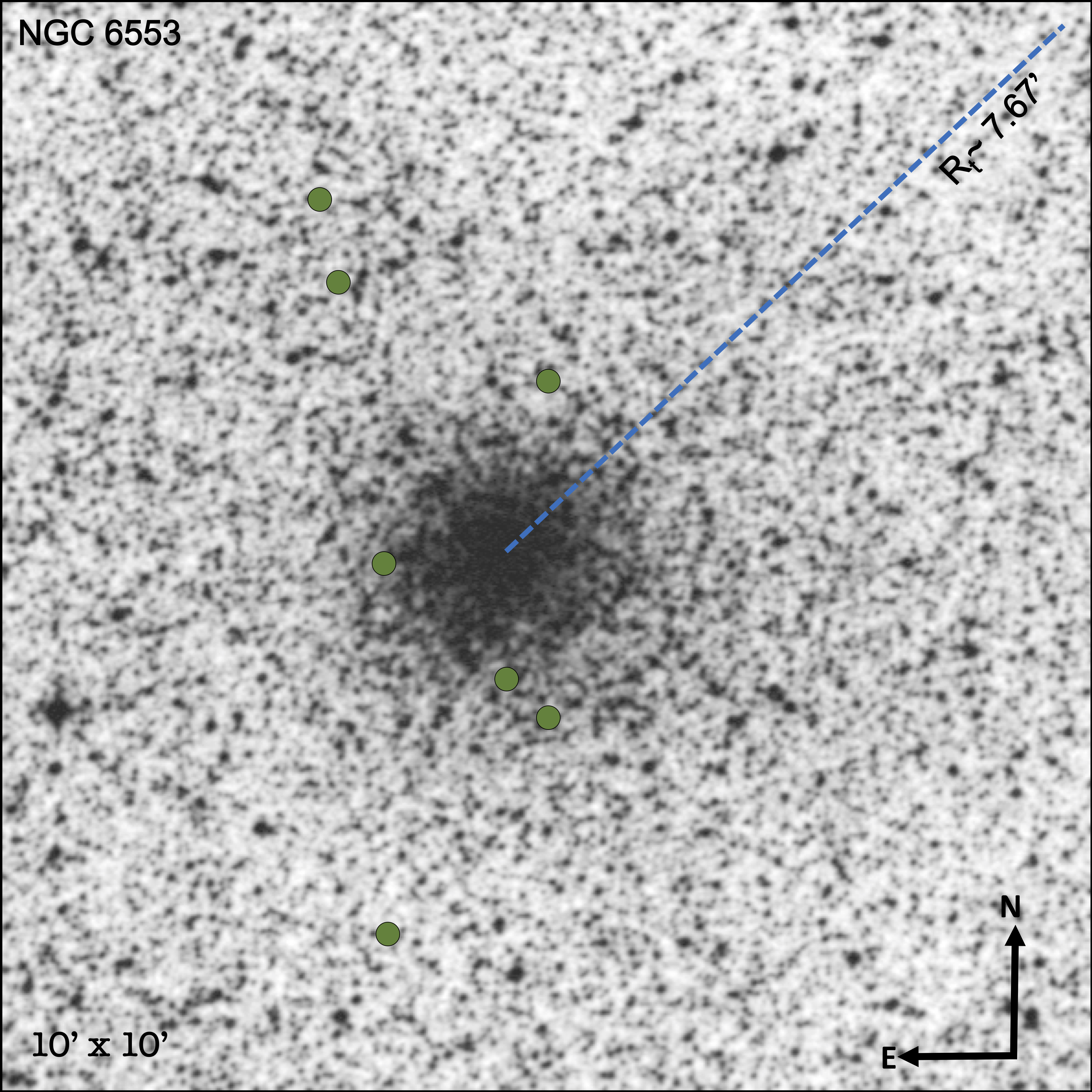}
  \caption{Distribution of the stars observed in NGC~6553 (green filled circles). The blue dashed line show the extent of  the tidal radius \citep[2010 edition]{harris96}.}
 \label{spatial}
 \end{figure}

\begin{figure}
\centering
  \includegraphics[width=3.5in,height=3.5in]{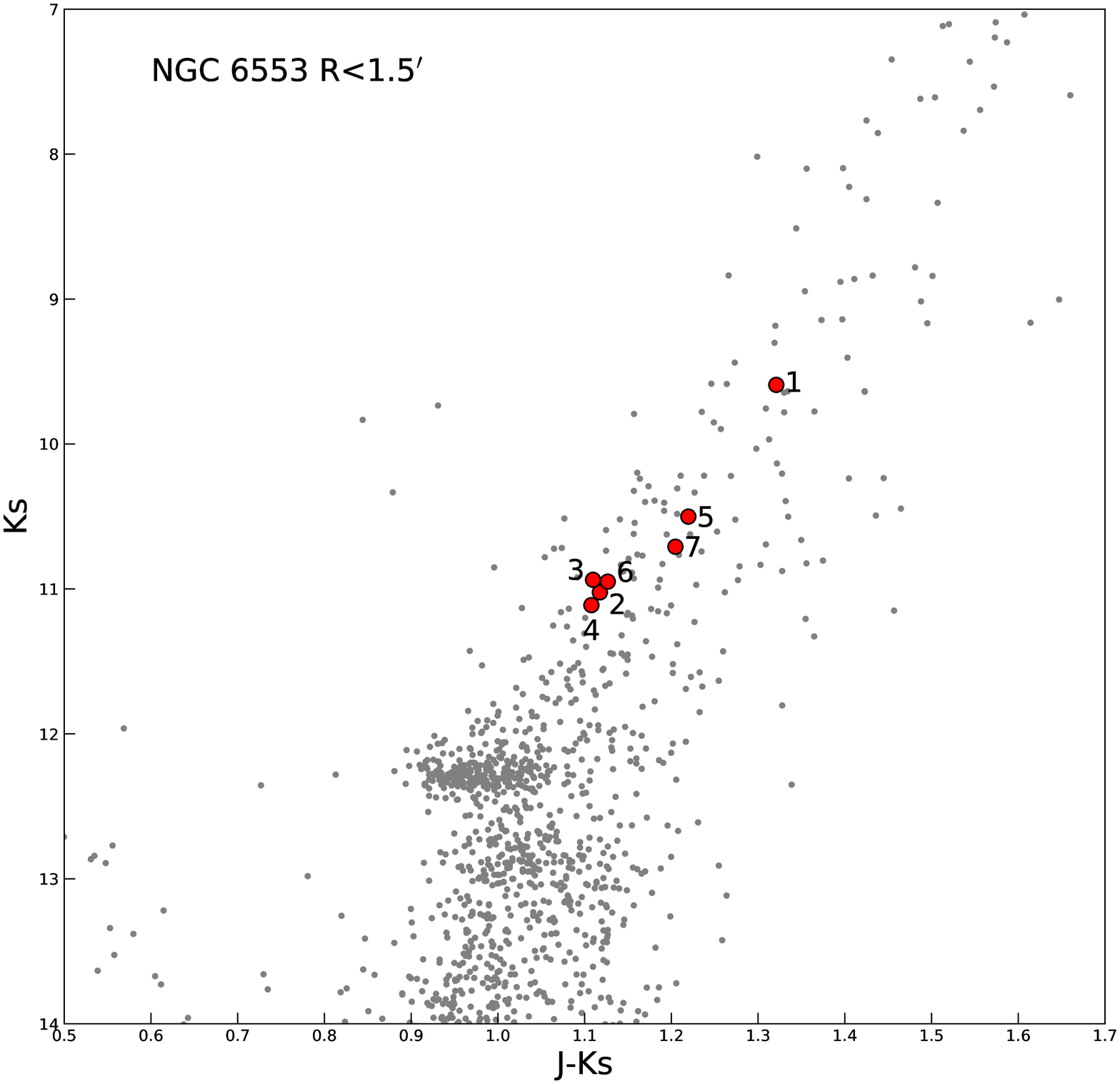}
  \caption{CMD of NGC~6553 from the VVV survey corrected by the VVV reddening maps \citep{gonzalez12}. The red filled circles represent our observed UVES sample.}
  \label{CMD}
 \end{figure}

The seven targets observed with FLAMES-UVES come from the membership list of NGC~6553 previously published in \citet{saviane12} and \citet{mauro14} using FORS2 Ca triplet spectroscopy and VVV photometry, whose spatial distribution is shown in Figure \ref{spatial}. All the stars of our sample belong to the upper  Red Giant Branch (RGB), as can be seen in the color-magnitude diagram (CMD) of the cluster (Figure \ref{CMD}). FLAMES-UVES data have a spectral resolution of about R$\simeq$47000. The data was taken with central wavelength 580\,nm, which covers the wavelength range 476-684\,nm. Our S/N is about 25 at 560\,nm (lower chip) and about 30 at 650\,nm (upper chip).


The reduction process includes  includes bias and flatfield corrections, wavelength calibration, spectral rectification, and sky subtraction. We apply the same procedure described in our previous articles \citep{munoz17,munoz18}.

The mean radial velocity for NGC~6553 in our sample is -3.86 $\pm$2.12 km $s^{-1}$, the  velocity dispersion is  5.62  km $s^{-1}$. This  radial velocity is compatible  with the values in the literature: \citet{saviane12} with four stars found a value of -9.0 $\pm$ 4.0  km $s^{-1}$, \citep[2010 edition]{harris96}  quotes a value of -3.2 $\pm$ 1.5  km $s^{-1}$ and \citet{tang17}  found a value of -0.14 $\pm$ 5.46 km $s^{-1}$.

Table \ref{param1}  lists the stellar parameters of our sample: ID, the J2000 coordinates (Ra and Dec), J, H, K$_{s}$ magnitudes from VVV PSF photometry, calibrated on the system of 2MASS \citep{mauro14,cohen17}, heliocentric radial velocity, Teff, log(g), micro-turbulent velocity ($v_{t}$) and metallicity.
Moreover,  Table \ref{iron-abun1} shows the metallicity values from \citet{saviane12}, \citet{mauro14} and \citet{tang17}. The procedure for the determination of the atmospheric parameters is discussed in the next section.

\begin{figure}
  \includegraphics[width=3.6in,height=3.6in]{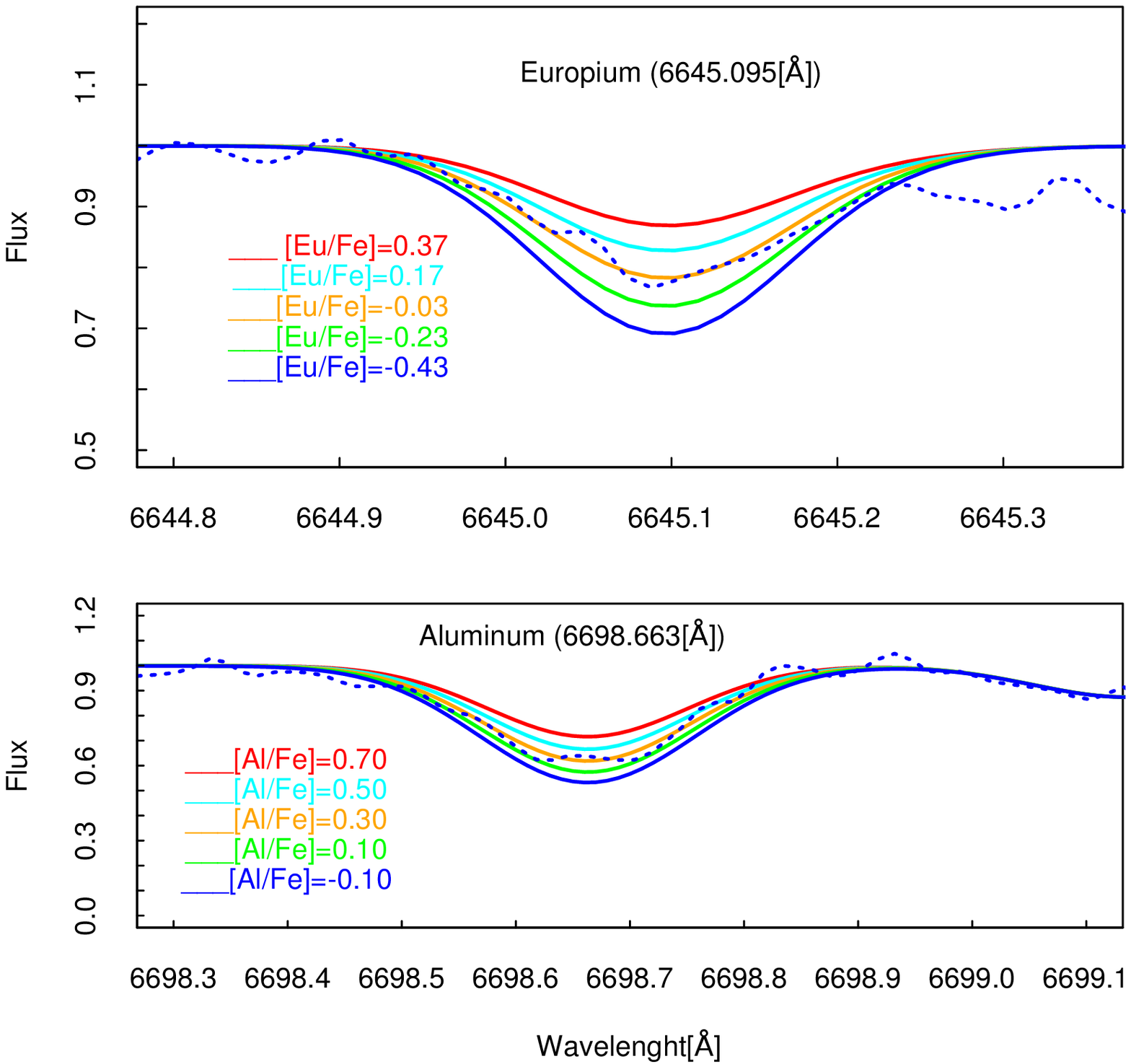}
  \caption{Spectrum synthesis fits for Europium (upper panel) and Aluminum (lower panel) lines for the star \#5 and \#4 respectively. The dashed line is the observed spectrum, and the solid color lines show the synthesized spectra corresponding to different abundances.}
  \label{synth}
 \end{figure}

\begin{figure}
  \includegraphics[width=3.2in,height=4.4in]{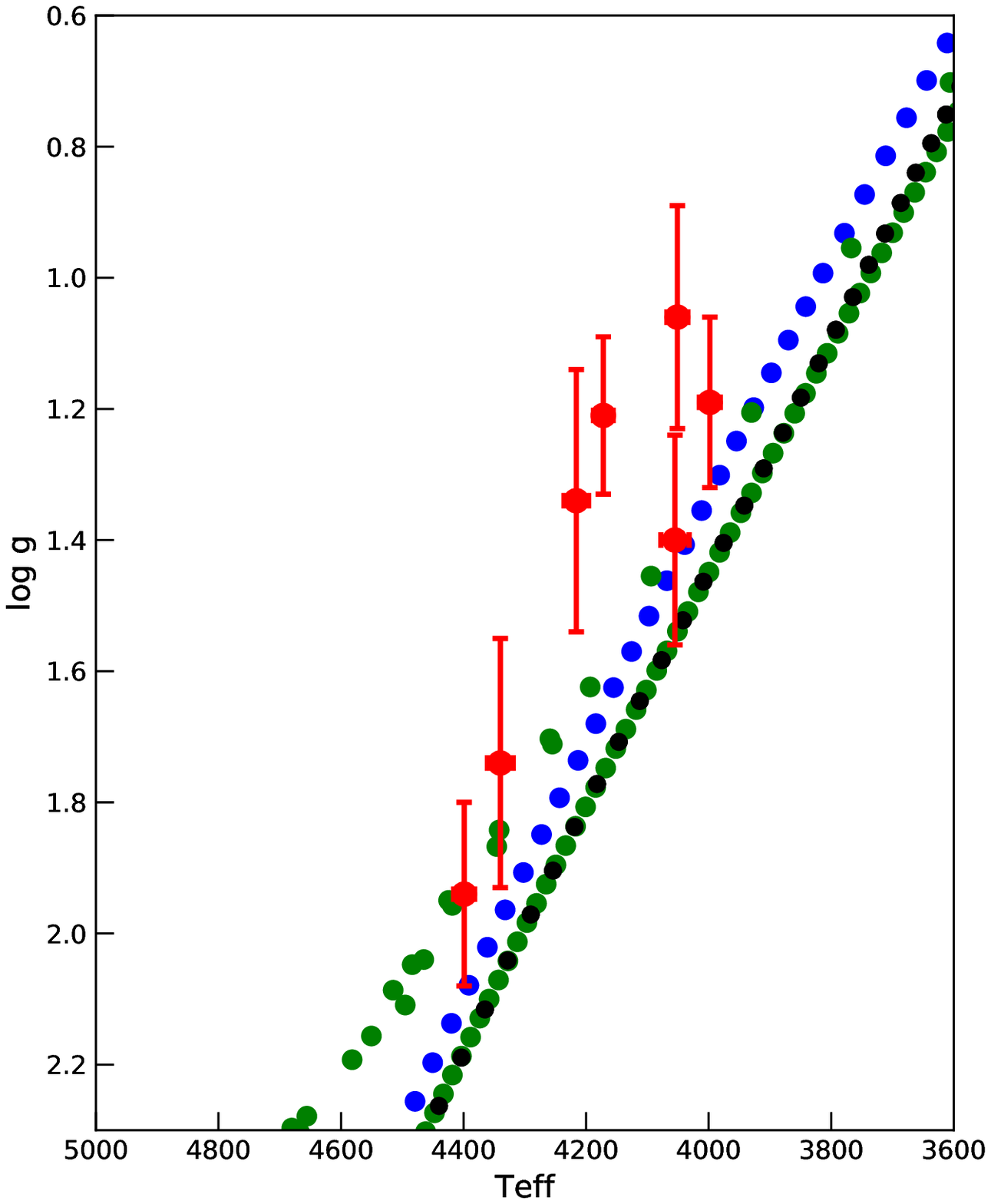}
  \caption{Log (g) vs $T_{eff}$ for NGC~6553. The black  points display an isochrone with  a metallicity of -0.16 dex, [$\alpha$/Fe]=+0.20 dex and age of 13 Gyr \citep[Dartmouth isochrone]{dotter08}. The green points is a   isochrone with  a metallicity of -0.16 dex and age of 13 Gyr \citep[PARSEC isochrone]{bressan12}.  The blue  points is a   isochrone with  a metallicity of -0.16 dex and age of 13 Gyr \citep[MESA isochrone]{choi16,dotter16}. }
  \label{CMD_v2}
 \end{figure}

\begin{table*}
\caption{Parameters of the observed stars for NGC~6553.}
\begin{threeparttable}[b]

\label{param1}
\begin{tabular}{ c c c c c c c c c c c c   }
\hline 
{\small{}ID} & {\small{}Ra} & {\small{}DEC } & {\small{}J } & {\small{}H } & {\small{}K$_{s}$} & {\small{}RV$_{H}$ } & {\small{}T$_{eff}$} & {\small{}log(g)} & {\small{}{[}Fe/H{]} } & $v_{t}$ & FeI/FeII \tabularnewline

 & {\small{}(h:m:s)} & {\small{}($\,^{\circ}{\rm }$:$^{\prime}$:$^{\prime\prime}$ )} & {\tiny{}(mag)} & {\tiny{}(mag)} & {\tiny{}(mag) } & {\tiny{}(km $s^{-1}$)} & {\small{}{[}K{]} } &  & dex & {\small{}{[}km/s{]} }  &    \tabularnewline
\hline 
1  & 18:09:15.66 &-25:56:00.77  & 10.86  & 9.90  & 9.63  & 6.12$\pm$0.21 & 4172$\pm$15 & 1.21$\pm$0.12 & -0.17$\pm$0.07 & 1.24$\pm$0.04 & 85/10 \tabularnewline
2  & 18:09:15.71 & -25:52:58.70 &12.17  & 11.276 & 11.035 &-8.07$\pm$0.19  & 3998$\pm$16  & 1.19$\pm$0.13  & -0.06$\pm$0.07 & 0.98$\pm$0.07 & 82/9 \tabularnewline
3  & 18:09:17.51 & -25:55:42.30  &12.05  & 11.17 & 10.94 &-7.49$\pm$0.41 &4216$\pm$19   &1.34$\pm$0.20  & -0.19$\pm$0.07 & 1.35$\pm$0.05 & 92/10  \tabularnewline
4  &18:09:22.39 & -25:54:37.94 &12.20  & 11.33 &11.10 & -10.95$\pm$0.27 & 4051$\pm$16 & 1.06$\pm$0.17& -0.07$\pm$0.07 & 0.89$\pm$0.10 &  88/11 \tabularnewline
5  &18:09:22.43& -25:57:59.32 &11.76  &10.80  & 10.52  &-1.81$\pm$0.42   & 4055$\pm$22 & 1.40$\pm$0.16&-0.16$\pm$0.07&1.05$\pm$0.12 &101/13 \tabularnewline

6  &18:09:23.98  &-25:52:01.20  & 12.07 & 11.22 &10.95 & -2.76$\pm$0.26& 4399$\pm$16  & 1.94$\pm$0.14 &-0.08$\pm$0.07  & 0.97$\pm$0.10 & 92/11 \tabularnewline
7  &18:09:24.67  &-25:51:11.10  &11.92  &11.03 & 10.71 &2.04$\pm$0.31 & 4340$\pm$20 & 1.74$\pm$0.19  & -0.22$\pm$0.07  & 1.47$\pm$0.06 &  81/8  \tabularnewline
\hline 

\end{tabular} 
Column 12: Numbers of line measured for FeI and FeII.
\end{threeparttable}

\end{table*}

\begin{table*}
\caption{Iron abundances from different authors for NGC~6553.}
\begin{threeparttable}[b]

\label {iron-abun1}

\centering
\small

\begin{tabular}{ c c c c c c }
\hline 
\hline
ID. & [Fe/H]$_{this\_work}$ & [Fe/H]$_{S12}$ &{[}Fe/H{]}$_{M14}$&{[}Fe/H{]}$_{D16}$&[Fe/H]$_{T17}$ \\
 &  &  & \\
\hline
1	&-0.17$\pm$0.07& -0.44& -0.27 $\pm$0.14&-0.15&- \\
2	&-0.06$\pm$0.07& 0.10&-0.02$\pm$0.14 &-0.13& -0.17\\
3	&-0.19$\pm$0.07&0.29& 0.30$\pm$0.14&-0.22& -\\
4	&-0.07$\pm$0.07& 0.12& -0.13$\pm$0.14 &-0.09&-0.16\\
5	&-0.16$\pm$0.07& 0.24&0.12 $\pm$0.14 &-0.13& -\\
6	&-0.08$\pm$0.07& -0.06&0.09 $\pm$0.14&-0.10&-0.08 \\
7	&-0.22$\pm$0.07& 0.00& 0.04$\pm$0.14&-0.14&-\\
\hline
\end{tabular}
\textbf{References.} S12: \citet{saviane12}; 
M14: \citet{mauro14}; \\ D16: \citet{dias16a};T17: \citet{tang17}.
\end{threeparttable}

\end{table*}

\section{Atmospheric Parameters and Abundances}

We have analyzed our sample of NGC~6553 stars using the local thermodynamic equilibrium (LTE) program MOOG \citep{sneden73}. Atmospheric models were performed using ATLAS9 \citep{kurucz70} and the line list for the chemical analysis is the same described in   \citet{villanova11} and widely used in several studies \citep{rain19,villanova13,munoz17,munoz18,mura17}. The stellar parameters Teff, vt, and log(g) were adjusted iteratively    and new stellar models  were calculated in an effort to remove trends in excitation potential and equivalent width vs. abundance for Teff and vt respectively, and to satisfy the ionization equilibrium for log(g). FeI and FeII were used for this latter objective. The [Fe/H] value of the model was changed at each iteration according to the output of the abundance analysis. Also, We present in Table \ref{param1} the uncertainties  for Teff, Log g and $v_{t}$,  these were estimated following \citet{gonzalezv98}  and \citet{neuforge97}.  The uncertainties for  $v_{t}$ were estimated  using the standard deviation in the slope of the least-squares fit of Abundance vs. reduced EW, the uncertainties for Teff were determined from  the uncertainty of the least-squares fit of abundance vs. excitation potential, in addition to the uncertainty  in the slope due to the uncertainties in $v_{t}$. Finally to  calculate  the uncertainty in logg, we include the contribution from the uncertainty in Teff in addition to the scatter in the Fe II line abundances. 

In Figure \ref{CMD_v2}  we found good agreement  among the stellar parameters derived in this study  and from three different model isochrones with similar metallicity and  with an age of 13 Gyr. The models used in this comparison are: PARSEC \citep{bressan12}, MESA \citep{choi18,dotter16} and  the Dartmouth \citep{dotter08} models. In Figure \ref{CMD_v2}  these models  are plotted  in green, blue and black respectively. Although we note a small offset among the data and the isochrone, it is important take account that an offset of $\sim$100 K between photometric and spectroscopic Teff  can arise because of uncertainties in the mixing length parameter and/or surface boundary condition \citep{choi18}. Also other studies find a similar mismatch especially associated with metal rich stars \citep{ness13}.


The reddening is high for most of  the bulge GCs, and  NGC~6553 is no exception. NGC~6553 has a color excess of E(B-V)=0.63  quoted by \citet[2010 edition]{harris96}  and  \citet{guarnieri98} found a  value of  E(B-V)=0.70.  The potentially high differential reddening and high crowding make it difficult to obtain the stellar parameters. In order to avoid the effect of the extinction and the differential reddening in the measurement of the stellar parameters, we decided to calculate the stellar parameters  directly from the spectra.

We used equivalent widths (EWs) of the spectral lines and the spectrum-synthesis method to obtain abundances of a large number of  elements, which are listed in Table \ref{abundances1}. We used the spectrum-synthesis method for lines affected by blending. In this case, we generated five synthetic spectra with different abundances for each line, and we estimated the best-fitting value as the one that minimises the rms scatter. Figure \ref{synth} shows an example of this method for two different lines. We carefully excluded the telluric contaminated lines in our analysis. The adopted solar abundances we use are reported in Table \ref{abundances1}. 


We carry out  an internal error analysis varying the stellar parameters ($T_{eff}$, log(g), [Fe/H], and vt) and redetermining abundances of star \#1, which  is  representative of our   whole sample (see Table \ref{error1}). Parameters were varied by $\Delta T_{eff}=+50$ K, $\Delta$log(g)=+0.13, $\Delta${[}Fe/H{]}=+0.03 dex, and $\Delta v_{t}=+0.09$ $km$ $s^{-1}$, which we estimated as our typical internal errors. The quantity  of variation of the parameter was calculated through three stars representative of our sample (\#1,\#2, and \#7) with relatively low, intermediate and high effective temperature respectively, according to the method   that was performed by  \citet{marino08}. The error introduced by the uncertainty on the EW ($\sigma_{S/N}$) was calculated by dividing the rms scatter by the square root of the number of the lines used for a given element and a given star.
For elements whose abundance was obtained by spectrum synthesis, the error is given in the output of the fitting procedure. The error for each [X/Fe] ratio as a result of uncertainties in atmospheric parameters and $\sigma_{S/N}$ are showed in Table \ref{error1}. The total internal error ($\sigma_{tot}$) is given by:


\begin{center}
\begin{equation}
\sigma_{tot}=\sqrt{\sigma^{2}_{T_{eff}}+\sigma^{2}_{log(g)}+\sigma^{2}_{v_{t}}+\sigma^{2}_{[Fe/H]}+\sigma^{2}_{S/N}}
\end{equation}
\end{center}


In Table \ref{error1} we compare the total internal error for all the elements measured with the observed error (standard deviation of the sample).

\begin{table*}
\caption{Abundances of the observed stars for NGC~6553.}
\begin{threeparttable}[b]
\label{abundances1}
\centering
\begin{tabular}{ l c  c c  c  c c  c  c c c  }
\hline 
\hline

El. & 1 & 2  & 3 & 4  & 5 & 6 & 7  & Cluster$^{1}$ &Sun \\


\hline 
$ [$O/Fe$] $   &   -0.17  &	-0.14  &	-0.05  &	-0.16 &	-0.06  &0.06   &	0.03     & -0.07$\pm$0.03   & 8.83 \\
&$\pm$0.03&$\pm$0.03&$\pm$0.05&$\pm$0.04&$\pm$0.05&$\pm$0.05&$\pm$0.04&\\

$ [$Na/Fe$]_{NLTE} $  &	 0.24 &	-0.09 &	0.47  &	0.22 &	0.60  &	-0.00     &	0.18  & +0.23$\pm$0.09   & 6.32\\
&$\pm$0.03&$\pm$0.05&$\pm$0.06&$\pm$0.05&$\pm$0.06&$\pm$0.06&$\pm$0.05&\\

$ [$Mg/Fe$] $  &	 0.27 &	0.23   &	0.22   &	0.17  &	0.07   &	0.14   &	0.29     & +0.20$\pm$0.03   & 7.56\\
&$\pm$0.04&$\pm$0.05&$\pm$0.04&$\pm$0.06&$\pm$0.05&$\pm$0.04&$\pm$0.05&\\

$ [$Al/Fe$] $  &	 0.28 &	0.13   &	0.34   &	0.34 &	0.53   &	0.21  &	0.40    & +0.32$\pm$0.05  & 6.43\\
&$\pm$ 0.03&$\pm$0.04&$\pm$0.04&$\pm$0.04&$\pm$0.04&$\pm$0.04&$\pm$0.04  \\

$ [$Si/Fe$] $  &	0.06  &	0.06   & 0.22   &	-0.29  &	-0.36   &-0.31   &	0.21   & -0.06$\pm$0.09   & 7.61\\
&$\pm$0.03&$\pm$0.05&$\pm$0.06&$\pm$0.05&$\pm$0.06&$\pm$0.05&$\pm$0.06&\\

$ [$Ca/Fe$] $  &	 -0.06   &	-0.15   &	-0.05   &	-0.04 &	0.15  &	0.04   &	-0.04     & -0.02$\pm$0.03   & 6.39\\
&$\pm$0.05&$\pm$0.06&$\pm$0.07&$\pm$0.07&$\pm$0.06&$\pm$0.07&$\pm$0.08&\\

$ [$Sc/Fe$] $  &	 -0.01  &	0.02  &	-0.16 &	-0.26 &	-0.09  &	-0.34   &	0.17    & +0.09$\pm$0.06   & 3.12\\
&$\pm$0.03&$\pm$0.05&$\pm$0.06&$\pm$0.05&$\pm$0.06&$\pm$0.05&$\pm$0.06&\\

$ [$Ti/Fe$] $  &0.17  &	0.20   &	0.24  &	0.31 &	0.56   &	0.43     &	0.31  & +0.32$\pm$0.05  & 4.94\\
&$\pm$0.06&$\pm$0.05&$\pm$0.06&$\pm$0.05&$\pm$0.07&$\pm$0.06&$\pm$0.06&\\

$ [$V/Fe$] $   &	0.33&	0.43  &	0.36  &	0.48 &	0.85  &	0.58  &	0.38  & +0.48$\pm$0.07   & 4.00\\
&$\pm$0.07&$\pm$0.05&$\pm$0.05&$\pm$0.07&$\pm$0.06&$\pm$0.07&$\pm$0.05&\\

$ [$Cr/Fe$] $   &	0.02&	-0.10  &	-0.09  &	-0.03 &	0.18  &	-0.02  &	0.03  & 0.00$\pm$0.03   & 5.63\\
&$\pm$0.09&$\pm$0.06&$\pm$0.09&$\pm$0.10&$\pm$0.09&$\pm$0.09&$\pm$0.05&\\

$ [$Mn/Fe$] $   &	-0.09&	-0.10  &	0.10  &	-0.27 &	-0.06  &	-0.37  &	-0.22  & -0.14$\pm$0.06   & 5.37\\
&$\pm$0.05&$\pm$0.07&$\pm$0.06&$\pm$0.07&$\pm$0.06&$\pm$0.06&$\pm$0.07&\\

$ [$Fe/H$] $   &	 -0.17&	-0.06 &	-0.19  &	-0.07 &	-0.16  &	-0.08  &	-0.22  & -0.14$\pm$0.02   & 7.50 \\
&$\pm$0.02&$\pm$0.03&$\pm$0.02&$\pm$0.02&$\pm$0.03&$\pm$0.03&$\pm$0.02&\\

$ [$Ni/Fe$] $  & 0.22  &	0.23  &	0.21  &	0.36 &	0.38 &	0.21  &	0.18  &+0.26$\pm$0.03  & 6.26\\
&$\pm$0.06&$\pm$0.06&$\pm$0.07&$\pm$0.08&$\pm$0.07&$\pm$0.08&$\pm$0.08&\\

$ [$Cu/Fe$] $  & 0.32  &	0.40  &	0.12  &	0.28 &	0.28 &	-0.17  &	0.41  &+0.23$\pm$0.08  & 4.19\\
&$\pm$0.07&$\pm$0.06&$\pm$0.08&$\pm$0.08&$\pm$0.09&$\pm$0.08&$\pm$0.08&\\

$ [$Y/Fe$] $  & 0.03 &    0.07  & 0.26  & 0.37 & 0.86 &0.17  & -0.03  &+0.25$\pm$0.11  & 2.25\\
&$\pm$0.06&$\pm$0.06&$\pm$0.07&$\pm$0.07&$\pm$0.06&$\pm$0.07&$\pm$0.06&\\

$ [$Zr/Fe$] $  & -0.52  &0.03  &-0.45  &-0.34 &	0.22 &	-0.70  & -0.50  &-0.32$\pm$0.16  & 2.56 \\
&$\pm$0.04&$\pm$0.05&$\pm$0.04&$\pm$0.08&$\pm$0.05&$\pm$0.05&$\pm$0.06&\\

$ [$Ba/Fe$] $  &-0.01  &	0.29 & -0.01 &  0.13  &	0.10  &	-0.40  & 0.18   & +0.04$\pm$0.08  & 2.34\\
&$\pm$0.05&$\pm$0.05&$\pm$0.06&$\pm$0.07&$\pm$0.06&$\pm$0.07&$\pm$0.05&\\

$ [$Ce/Fe$] $  &-0.24 &	-0.07 & -0.10 &  -0.12  &	0.03  &	-0.31  & -0.08   & -0.13$\pm$0.04  & 1.53\\
&$\pm$0.08&$\pm$0.04&$\pm$0.04&$\pm$0.09&$\pm$0.08&$\pm$0.08&$\pm$0.10&\\

$ [$Nd/Fe$] $  &-0.29 &	-0.01 & -0.38 &  -0.41  &	-0.19  &	-0.76  & -0.37   & -0.34$\pm$0.09  & 1.59\\
&$\pm$0.06&$\pm$0.08&$\pm$0.06&$\pm$0.07&$\pm$0.10&$\pm$0.08&$\pm$0.07&\\

$ [$Eu/Fe$] $  &	-0.06  &	0.06  &	0.08  & -0.03  &	-0.01  &	-0.23  &	0.02   & -0.02$\pm$0.04 & 0.52\\
&$\pm$0.04&$\pm$0.03&$\pm$0.04&$\pm$0.05&$\pm$0.04&$\pm$0.04&$\pm$0.05&\\

\hline

\end{tabular}
Columns 2-8: abundances of the observed stars. Column 9: mean abundance for the cluster. Column 10: abundances adopted for the Sun in this paper. Abundances for the Sun are indicated as log$\epsilon$(El.).
The errors presented for each abundance was calculated by dividing the rms scatter by the square root of the number of the lines used for a given element and a given star. For elements whose abundance was obtained by
spectrum-synthesis, the error is the output of the fitting procedure.\\
(1) The errors are the statistical errors obtained of the mean.\\
\end{threeparttable}
\end{table*}

\begin{table*}
\caption{Estimated errors on abundances for NGC~6553, due to errors on atmospherics parameters and to spectral noise, compared with the observed errors.}
\label{error1}
\centering
\begin{tabular}{ l c  c c  c  c c  c  c c  c }
\hline 
\hline
	ID   &  $\Delta T_{eff} =50 K $ & $ \Delta log(g)=0.13$  & $\Delta v_{t}= 0.09$	& $ \Delta [Fe/H]=0.03 $ & $\sigma_{S/N}$ & $\sigma_{tot}$  & $\sigma_{obs}$\\		
	\hline					
	$ \Delta ([O/Fe]) $     &	0.01 & -0.02 	   &	0.06   &	0.02   &	0.03  &	0.07 &	0.09\\ 
	$ \Delta ([Na/Fe]) $  &	-0.08 &	-0.04  &	0.00   &	-0.06   &	0.03  &	0.11 &	0.24\\ 	
	
	$ \Delta ([Mg/Fe]) $    &	-0.07  &	0.00   &	0.03   &	0.01   &	0.04  &	0.09  &	0.08\\ 	
	$ \Delta ([Al/Fe]) $    &	-0.01  &	0.01  &	    0.05  &	    0.03  &	0.03  &	0.07  &	0.13\\ 	
	
	$ \Delta ([Si/Fe]) $    &	0.07  &	-0.07  &	-0.08   &	0.06   &	0.05  &	0.15  &	0.25\\ 
			
	$ \Delta ([Ca/Fe]) $    &	-0.07   &	-0.01  &	0.03  &	-0.02  &	0.05  &	0.09 &	0.09\\

	$ \Delta ([Sc/Fe]) $    &	0.04   &	-0.04  &	-0.01  &	0.00   &	0.03&	0.06  &	0.17\\ 	
	
	$ \Delta ([Ti/Fe]) $  &	-0.06    &	0.00 &	-0.08   &	-0.01  &	0.06 &	0.11  &	0.14\\ 	
    
		$ \Delta ([V/Fe]) $    &-0.06  &	0.00  &	-0.10  &	-0.01   &	0.07&	0.14  &	0.18\\ 	
     	$ \Delta ([Cr/Fe]) $    &	-0.09  &	-0.05   &	-0.01   &	-0.04   &	0.09  &	0.15  &	0.09\\ 
  	  $ \Delta ([Mn/Fe]) $    &	0.10  &	0.12   &	0.10   &	0.04   &	0.05  &	0.19 &	0.16\\ 	
 
	$ \Delta ([Fe/H]) $     &	0.01   &	0.02   &	-0.06   &	0.00   &	0.02  &	0.07  &	0.06\\ 	

	$ \Delta ([Ni/Fe]) $  &	0.12  &	0.11   &	0.15   &	0.13   &	0.06  &	0.26 &	0.08\\
    	$ \Delta ([Cu/Fe]) $    &	0.06  &	0.10  &	0.09   &	0.12   &	0.07  &	0.20  &	0.20\\ 
    	 
         $ \Delta ([Y/Fe]) $    &	0.12 &	0.13   &	0.16   &	0.16   &	0.06  &	0.29  &	0.30\\ 

        $ \Delta ([Zr/Fe]) $    &	0.05  &	0.06   &	0.11   &	0.07   &	0.04  &	0.16  &	0.33\\ 	
	$ \Delta ([Ba/Fe]) $    &	0.1  &	0.07   &	0.15   &	0.11   &	0.05  &	0.23  &	0.22\\ 	
	
    $ \Delta ([Ce/Fe]) $    &0.13  &	0.08   &	0.13   &	0.10   &	0.08  &	0.24&	0.11\\ 	
	$ \Delta ([Nd/Fe]) $    &	0.06  &	0.05   &	0.10   &	0.08   &	0.06  &	0.15  &	0.23\\

    $ \Delta ([Eu/Fe]) $    &	-0.01  &	0.01   &	0.09   &	-0.03   &	0.04  &	0.10 &	0.10\\

\hline

\end{tabular}
\end{table*}

\section {Results}
In this section, we will discuss and examine in detail our results. Furthermore, we compare them with the literature, focusing  on our previous articles \citep{munoz17,munoz18}, which present an identical analysis of two bulge GCs (NGC~6440 and NGC~6528).

\subsection{Iron} 

We found a mean [Fe/H] value for the cluster of [Fe/H]=$-0.14\pm0.07$ dex. The scatter observed in this cluster is $\sigma_{obs}$=0.06, which is consistent  with the  total expected observational error  of $\sigma_{tot}$=0.07, indicating a homogeneous iron content. \citet{saviane12} using CaII triplet  found a mean in iron of [Fe/H]=-0.16$\pm$0.06 in their sample. We present in table  \ref{iron-abun1} the targets which we have in common.   \citet{mauro14} found a mean  value in iron of [Fe/H]=0.02 using the CaII triplet equivalent widths from \citet{saviane12} but using near IR  instead of optical photometry for the analysis (see Table \ref{iron-abun1}).  These studies are compatible with our results taking into account the uncertainties. It is interesting to note that \citet{saviane12} found a scatter of $\sigma$=0.06, compatible with no spread in iron in  agreement with our finding of homogeneity in iron. \citet{mauro14} found a larger scatter of  $\sigma$=0.17. However, this value is similar to their errors, indicating homogeneity  in metallicity, in agreement with  our results.

\citet{tang17} using high-resolution spectra in a sample of ten  members of NGC~6553 from APOGEE found a mean of [Fe/H]=-0.15$\pm$0.05 with a spread in iron of $\sigma$=0.05, in excellent agreement with our sample.  We have three stars in common with Tang (Table \ref{iron-abun1}) - their results for these stars are compatible with our results  taking into account the errors.

\citet{dias16a} studied  the low-resolution optical spectra  of the same stars from \citet{saviane12}. They used a full-spectrum fitting technique to derive the abundance. They found an average metallicity for NGC~6553 of [Fe/H]=-0.13$\pm$0.01 in agree with our finding. Also, we found good accordance star by star (see Table \ref{iron-abun1}).

Finally, \citet{ernandes18} study the iron peak-elements in four stars member of NGC~6553 using high-resolution spectroscopy. They found a metallicity of [Fe/H]=-0.20 dex with a scatter of  $\sigma$=0.02, in good agreement  again with our results.

\begin{figure}
\centering
  \includegraphics[width=3.5in,height=7in]{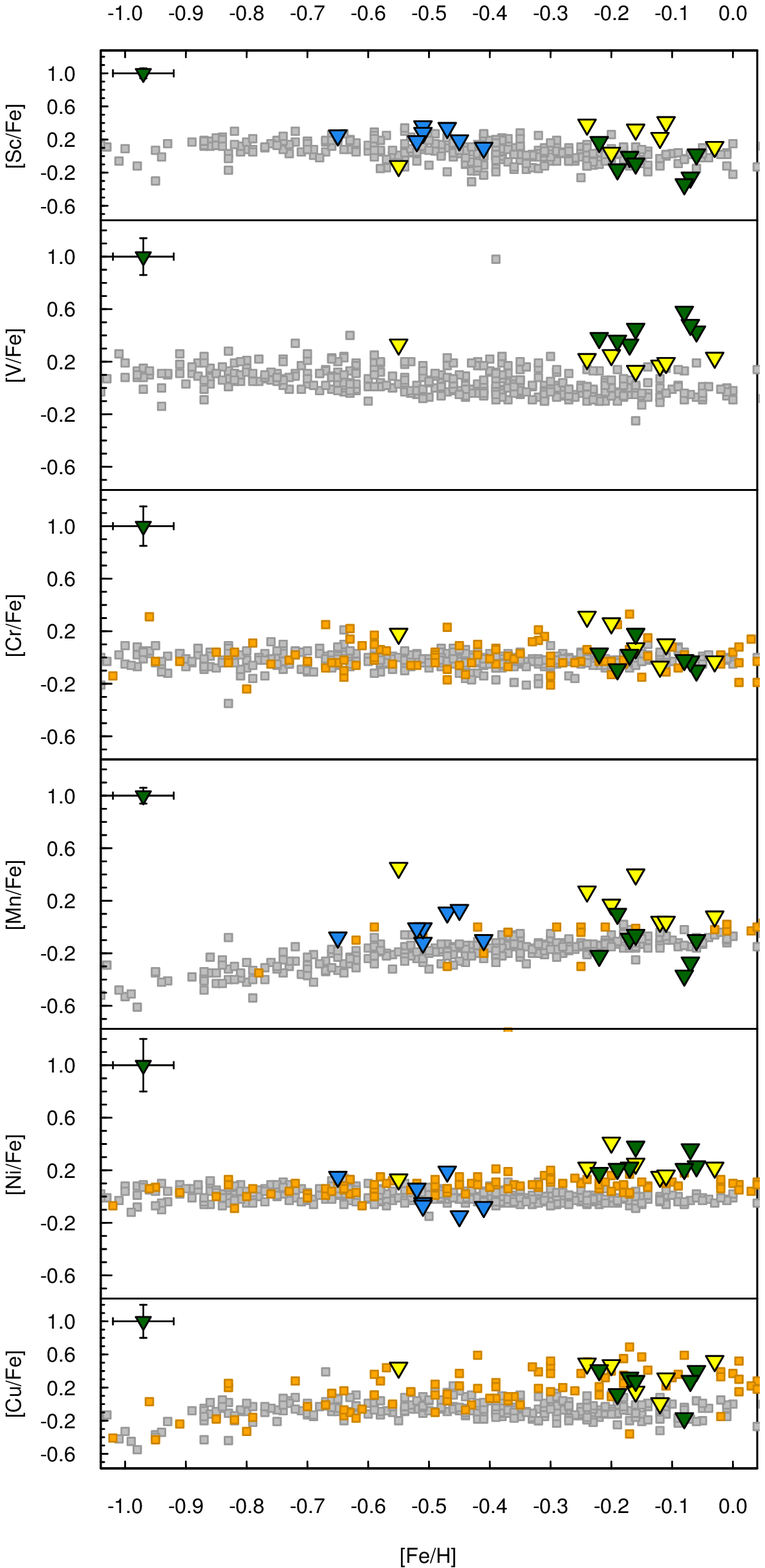}
  \caption{[Sc,V,Cr,Mn,Ni,Cu/Fe] vs [Fe/H]. Filled dark green triangles  are our data for NGC~6553 (this study), Filled yellow triangles  are our data for NGC~6528 \citep{munoz18}, filled blue triangles: NGC~6440 \citep{munoz17}, filled orange squares: Bulge field stars \citep{barbuy13,johnson14}, filled gray squares: halo and disk  stars \citep{fulbright00,francois07,reddy03,reddy06}.}
  \label{iron-ele}
 \end{figure}
\subsection{Iron-peak elements}

We have measured the abundance of seven iron-peak elements: Sc, V, Cr, Mn, Fe, Ni, and Cu (see Table \ref{abundances1} and Figure \ref{iron-ele}). We have analyzed the iron in detail in the previous section.

In Figure \ref{iron-ele} we plotted iron-peak elements versus [Fe/H]  comparing with our previous studies for NGC~6440 \citep{munoz17} and NGC~6528 \citep{munoz18}. We found good agreement with NGC~6528 for the case of Cr, Ni, and Cu. All of  these elements are super-solar except for Mn. Vanadium shows a very high super-solar abundance; however, the observational error is quite
quite large for this element.

Three of the iron-peak elements (Cr, Mn, Ni) were analysed in APOGEE DR13.
The mean values of the ten stars presented in \citet{tang17}  are: [Cr/Fe]=0.00, [Mn/Fe]=0.04, [Ni/Fe]=0.06. Cr shows the same mean value found in this study as in \citet{tang17}.  V and Cu given by APOGEE DR13 are subject to large uncertainties \citep{tang17}, therefore, we made no comparison for V and Cu.

 \citet{ernandes18} studied some iron-peak elements (Sc, V, Mn, Cu, and Zn) in NGC~6553.  Their results are compatible with our results for Sc and Cu, taking into account the uncertainties.  The more substantial  difference is for Vanadium, although our error for this element is large.

Similar to the case of NGC~6528 and NGC~6440, the super solar abundance for most of the iron-peak elements is evidence of early pollution by SN explosion(s).

\subsection{$\alpha$ elements}
\label{alpha}
Alpha elements are suggested to come from SN II explosions
at an early epoch. We managed to measure five $\alpha$ elements (O, Mg, Si, Ca and Ti).  NGC~6553 shows very similar behavior to NGC~6528 for the case of the alpha-elements, with a very strong overabundance relative to the solar scale for Mg and Ti and with solar abundance for O, Si and Ca.

Using the alpha-elements Mg, Si, Ca and Ti to obtain the mean, we obtain [$\alpha$/Fe]=$0.11\pm0.05$.

All the bulge GCs of our studies (NGC~6440, NGC~6528 and NGC~6553), and including NGC~6441 (Gratton et al. 2006, 2007) and HP-1 (Barbuy et al. 2016), show good agreement with the trend of the bulge (See figure 6 and 7) for the alpha elements in general. Although, in the case of NGC~6553, this one shows some compatibility with the bulge trend as well with the disk trend (see Figure \ref{alpha1}).

 We did not find a clear Si spread in NGC 6553 ($\sigma_{tot}$=0.15,$\sigma_{obs}$=0.25), similar to the case of \citet{tang18}. It is  interesting to note that in NGC~6528 \citep{munoz18} we found a similar behavior ($\sigma_{tot}$=0.11,$\sigma_{obs}$=0.14). An intrinsic spread in Silicon is mainly  found in metal-poor GCs or massive GCs \citep{tang18,ventura12,dantona16}. Therefore, we did not expect to find a  spread in Silicon in NGC~6553 or NGC~6528. However, the elements for  which we  did expect to find a significant spread, viz. O and Mg or both, in fact show little or no scatter, basically equal to the total error (Table \ref{error1}). We will discuss  O and Mg in the next sections.

     \begin{figure}
\centering
  \includegraphics[width=3.5in,height=6.4in]{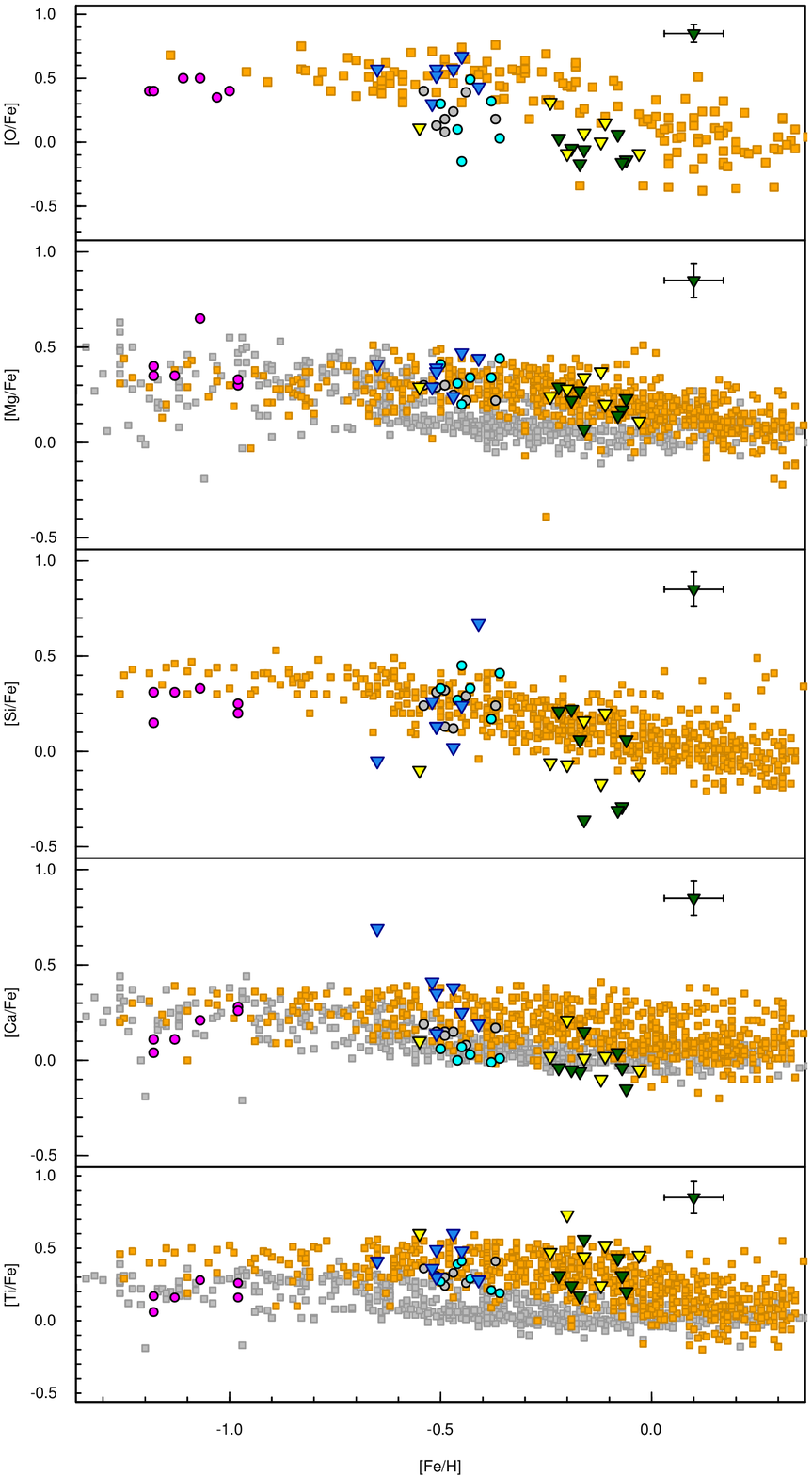}
  \caption{$[$O/Fe$]$,[Mg/Fe], [Si/Fe], [Ca/Fe], [Ti/Fe] vs [Fe/H]. Filled dark green triangles  are our data for NGC~6553 (this study), filled yellow triangles are our data for NGC 6528 \citep{munoz18}, filled blue triangles: NGC 6440 \citep{munoz17},filled cyan circles :  NGC6441 \citep{gratton06,gratton07}, filled gray  circless: NGC~5927\citep{mura17}, filled magenta triangles: HP1 \citep{barbuy16}, filled orange squares: bulge field stars\citep{gonzalez12},  filled grey squares : Halo and Disk fields stars \citep{venn04}.}
  \label{alpha2}
 \end{figure}
  
\begin{figure}
\centering
  \includegraphics[width=3.3in,height=3.3in]{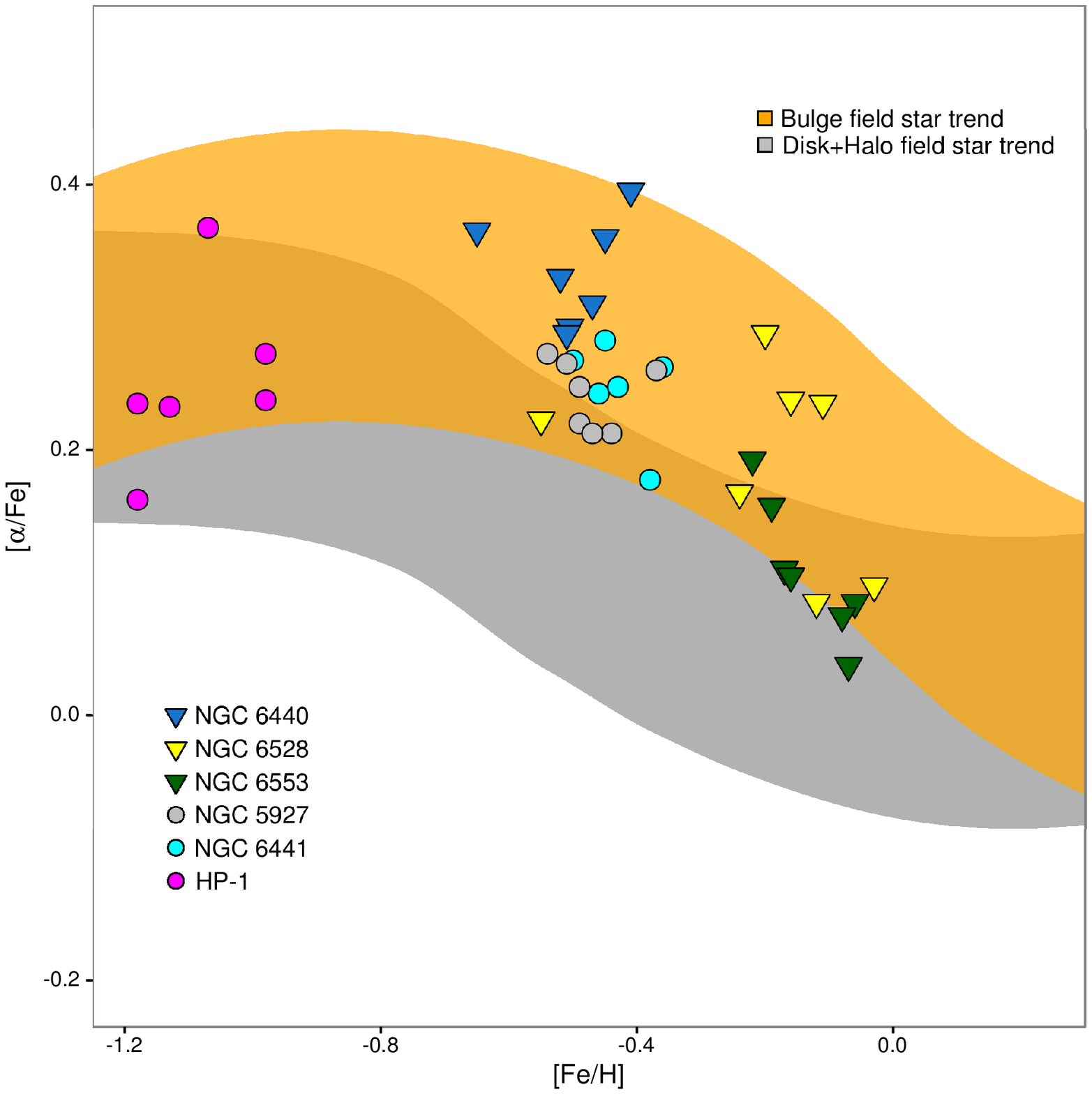}
  \caption{[alpha/Fe] vs [Fe/H]. Filled  dark green triangles are our data for NGC~6553 (this study). Filled yellow triangles: NGC~6528, filled blue triangles: NGC~6440 \citep{munoz17}, filled cyan circles :  NGC6441 \citep{gratton06,gratton07}, filled gray  circless: NGC~5927\citep{mura17}, filled magenta triangles: HP1 \citep{barbuy16}, filled orange squares: bulge field stars\citep{gonzalez12},   filled grey squares : Halo and Disk fields stars \citep{venn04}.}
  \label{alpha1}
 \end{figure}

\subsection{Na-O anticorrelation}

Without a doubt, the Na-O anticorrelation has given us a powerful tool to study the MPs in GCs. Currently, virtually all old massive globular clusters clearly show this remarkable anticorrelation with at least one clear exception - Ruprecht~106 \citep{villanova13,dotter18}. However, it has been established that the extension of this anticorrelation is mainly connected with the mass and metallicity of the GC \citep{carretta09b,carretta15,carretta11,carretta10a}.

\begin{figure}
  \centering
  \includegraphics[width=3.3in,height=8.4in]{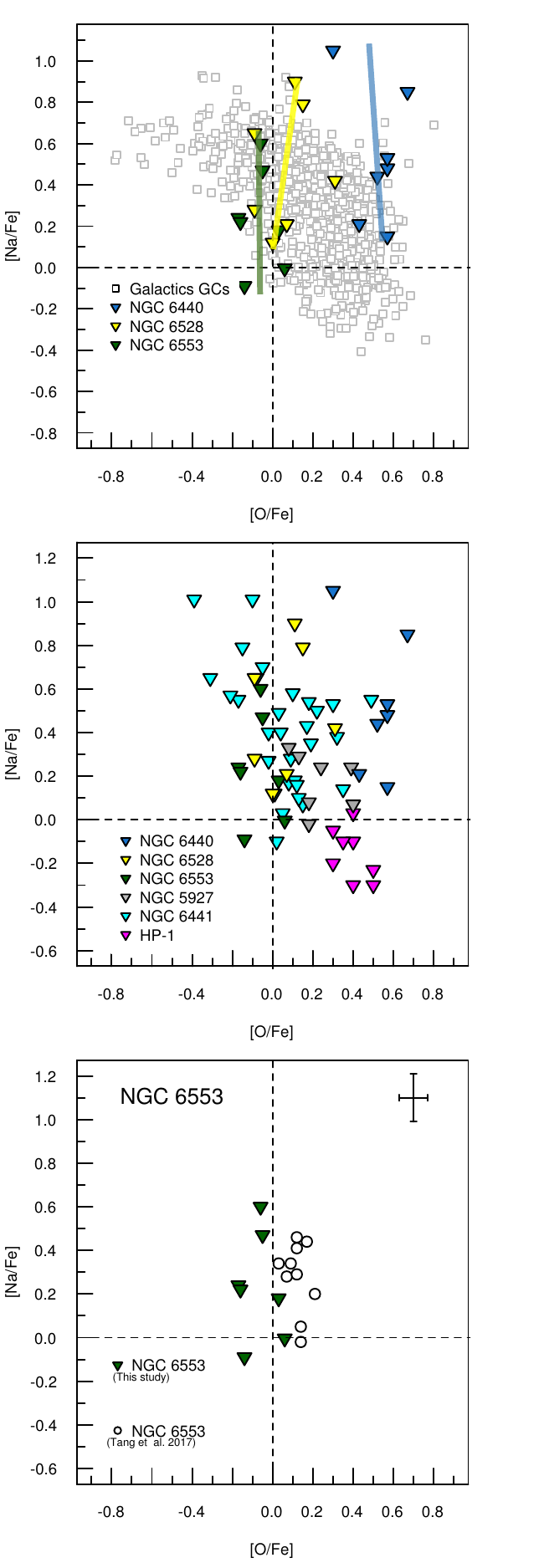}
  \caption{[O/Fe] vs [Na/Fe].
  Filled  dark green  triangles are our data for NGC~6553, filled yellow triangles: NGC~6528 \citep{munoz18}, filled blue triangles: NGC~6440 \citep{munoz17}, filled gray triangles : NGC~5927\citep{mura17}, filled cyan  triangles :  NGC~6441 \citep{gratton06,gratton07}, filled magenta triangles: HP1 \citep{barbuy16},  filled grey square: Galactic GCs from \citet{carretta09b}, open circles : NGC~6553 \citep{tang17}. The  solid vertical line represents the trend in each GC. }
    \label{onafig}

\end{figure}

In our previous articles \citep{munoz17,munoz18}, we have found in NGC~6528 and NGC~6440 a peculiar O-Na anti-correlation, basically vertical and with a very short, if any, horizontal extension, implying a Na but no significant O spread. NGC~6553 follows this pattern, with a very  low scatter in O of $\sigma_{obs}$=0.09 (compared to a total expected error of 0.07) and a more significant spread in Na of $\sigma_{obs}$=0.24 (compared to an expected error of 0.11). A similar pattern was shown in Tang et al. (2017): a small scatter in  Oxygen ($\sigma_{obs}$=0.05), and a spread  much greater in Na ($\sigma_{obs}$=0.15) in comparison with their  expected error. Their scatter values are very close to our results (see Figure \ref{onafig}). 

Our results are in agreement with previous findings about the mentioned in various articles about the extension of the anti-correlation Na-O and its dependence on cluster mass. The GCs from our previous studies (i.e., NGC~6440 and  NGC~6528) including the one presented  here - NGC~6553, have masses between (2.35$\pm$0.19)x10$^{5}$ and (8.96$\pm$1.85)x10$^{5}$M$_{\odot}$ \citep{baumgardt18}. Basically, these are intermediate mass GCs in comparison with other Galactic GCs. We find definite evidence that such intermediate mass bulge GCs have at most a short, almost vertical Na-O anticorrelation, extension, without a significant spread in Oxygen. On the other hand, contrasting its results with NGC~6441 \citep{gratton06,gratton07}, a massive Bulge GCs of (1.23$\pm$0.01)x10$^{6}$M$_{\odot}$\citep{baumgardt18}, we noticed a broader extension of the correlation. Finally, comparing with the GC HP~1 with a low mass of (1.11$\pm$0.38)x10$^{5}$M$_{\odot}$\citep{baumgardt18}, which have an unclear O-Na anti-correlation \citep{barbuy16}, although HP~1 is the most metal poor  among the GCs compared in this research, with a metallicity of [Fe/H]=-1.06$\pm$0.10 dex \citep{barbuy16}. Therefore, apparently  there is a   relationship between the mass and metallicity  of the bulge GCs  with their Na-O extension, in agreement with the literature.

\subsection{Mg-Al and  Na-Al}
 The study of the relationship between Mg and Al is another  useful tool when studying MPs in  GCs. Many authors have found an anti-correlation between these two elements \citep{carretta09b,meszaros15}, but unlike the  Na-O anticorrelation, this is present in fewer GCs studied so far. Nevertheless, similar to the case of Na-O, the extension of this anti-correlation strongly depends on mass and metallicity \citep{pancino17}.

\begin{figure*}
\centering
  \includegraphics[width=7.1in,height=4.2in]{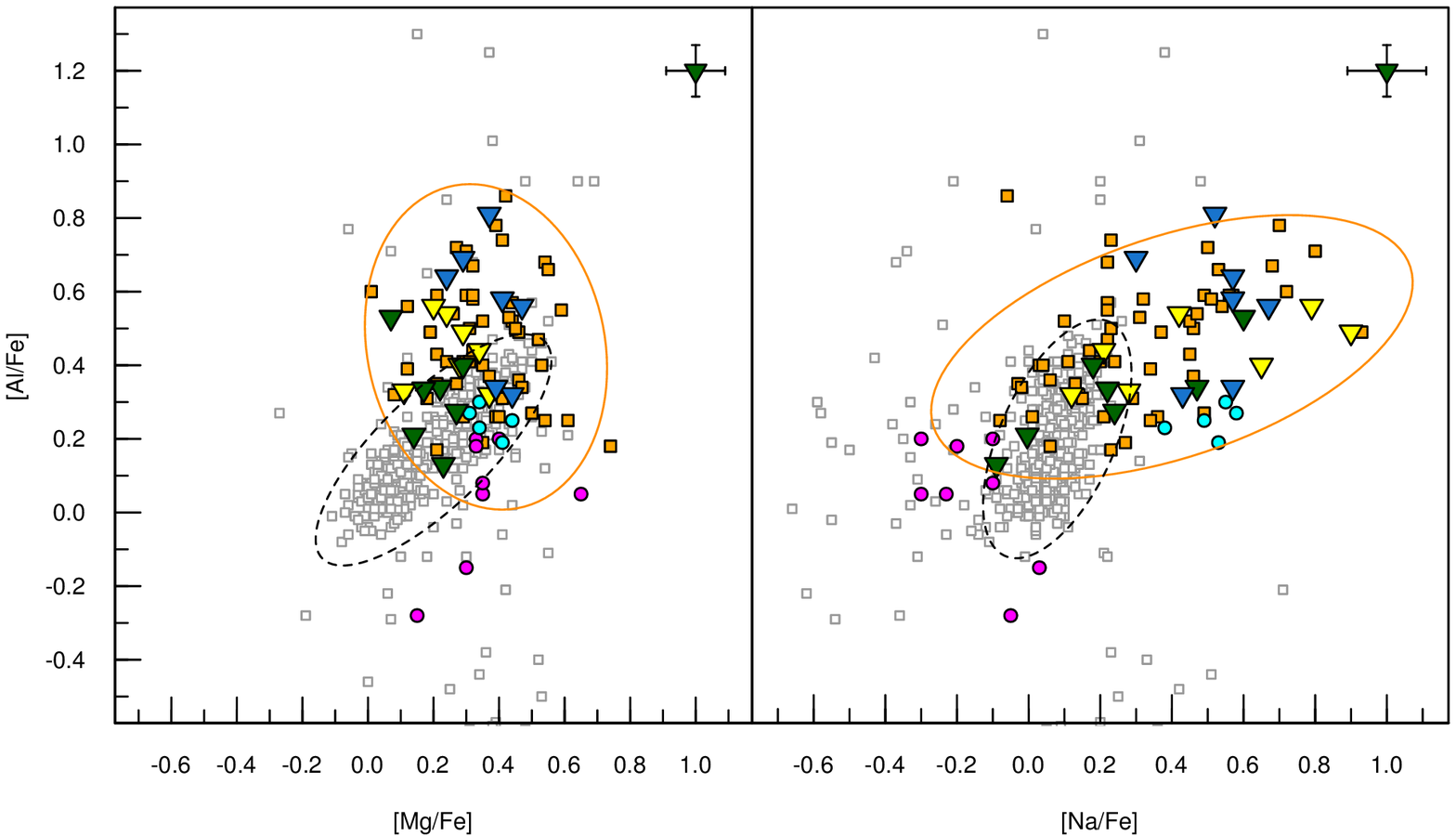}
  \caption{[Mg/Fe] vs [Al/Fe] and [Na/Fe] vs [Al/Fe] . Filled  dark green  triangles are our data for NGC~6553, filled yellow  triangles: NGC~6528 \citep{munoz18}, filled blue triangles: NGC~6440 (Munoz et al. 2017), filled cyan  triangles :  NGC~6441 \citep{gratton06,gratton07}, filled magenta triangles: HP1 \citep{barbuy16}, filled orange squares: bulge field stars \citep{lecureur07}, filled grey squares : Halo and Disk fields stars \citep{fulbright00,reddy03,reddy06,barklem05,cayrel04}.}
  \label{mg-al-na}
 \end{figure*}

Similar to the case of NGC~6528 from our previous study, we have not found a Mg-Al anti-correlation in this cluster (see Figure \ref{mg-al-na}). The spread of Mg is basically the same as the total error ($\sigma_{tot}$=0.09; $\sigma_{obs}$=0.08) and the spread of Al is somewhat greater than the total error ($\sigma_{tot}$=0.07; $\sigma_{obs}$=0.13). These results are in agreement with the result presented by \citet{tang17}. They found a mean of Mg of [Mg/Al]=0.15 with a scatter of $\sigma$=0.02, while in the case of Al they  found a mean of [Al/Fe]=0.20 with a scatter of ($\sigma$=0.14). 

We build a plot using  Na and Al (see Figure \ref{mg-al-na}), the light elements showing the most significant  spread in our sample.  We found a  good agreement with bulge field star trend with an important extension.  The spread in these elements allows us to verify  the presence of MPs in this GC. Also, again we found a good concordance  with the bulge GC NGC~6528. Finally, there are regions in this diagram where it is possible  to disentangle bulge and disk stars regardless if they are in clusters or the field (see figure \ref{mg-al-na}).

   \begin{figure}[htb]
  \includegraphics[width=3.5in,height=7in]{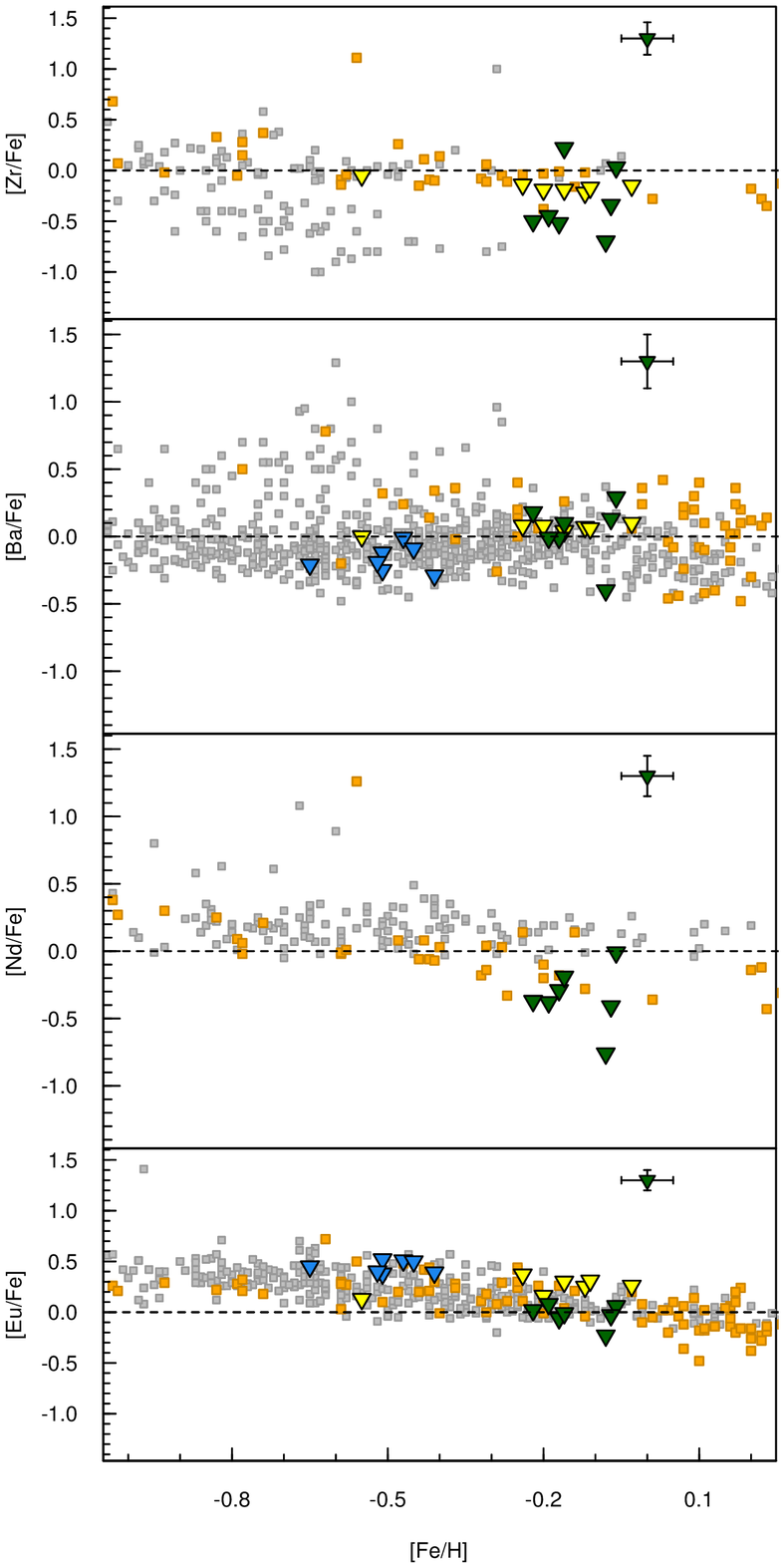}
  \caption{[Eu, Ba/Fe] vs [Fe/H].  Filled  dark green  triangles are our data for NGC~6553, filled yellow  triangles: NGC~6528 \citep{munoz18}, filled blue triangles: NGC~6440 \citep{munoz17}, filled orange squares: bulge field stars \citep{vanderswaelmen16}, filled  grey squares: Halo and disk stars \citep{fulbright00,francois07,reddy06,barklem05,venn04}.}

  \label{heavy}
 \end{figure}

\begin{figure}[htb]
 \centering
  \includegraphics[width=3.5in,height=3.7in]{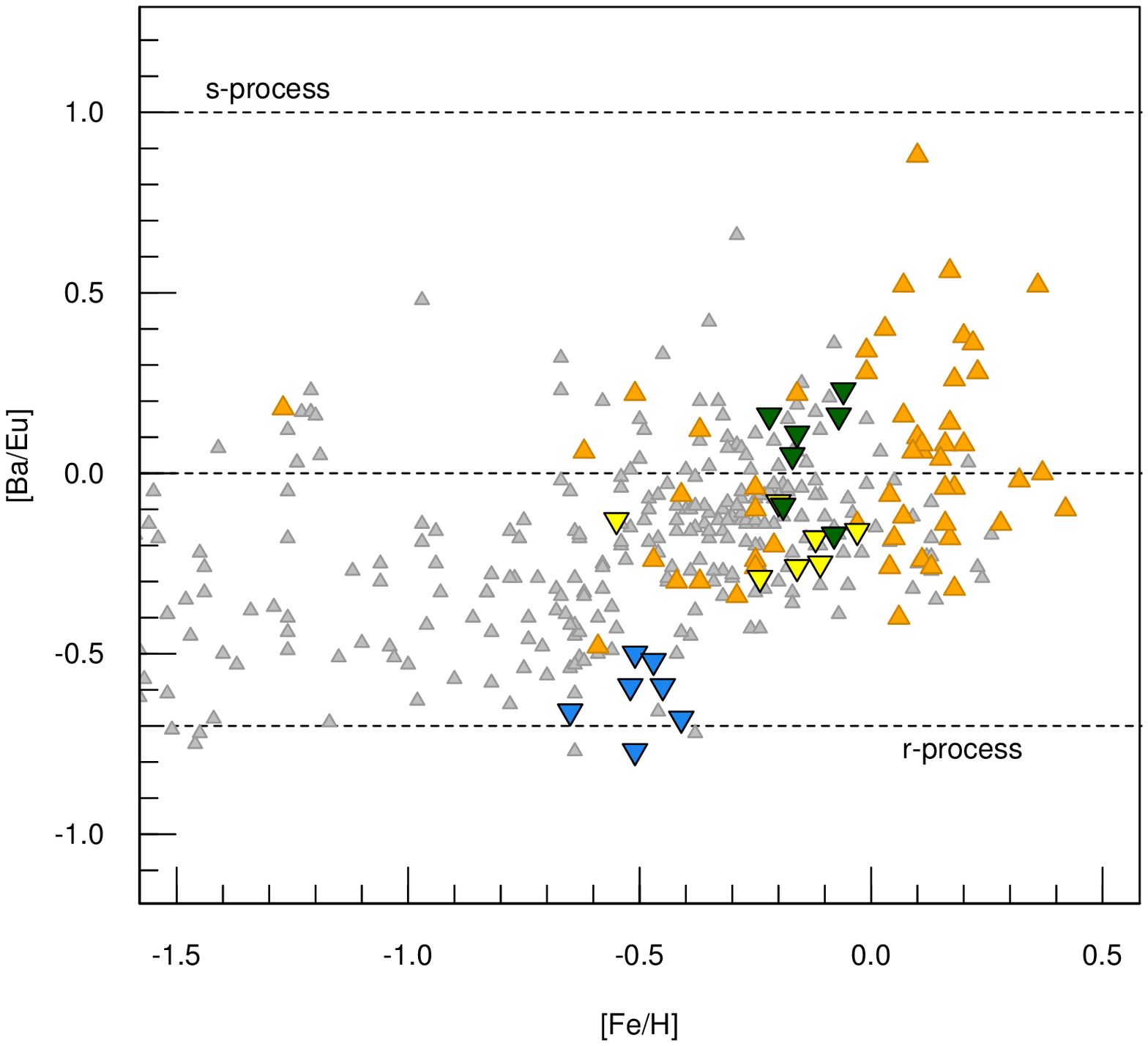}
  \caption{[Ba/Eu] vs [Fe/H]. Filled  dark green  triangles are our data for NGC~6553, filled yellow  triangles: NGC~6528 \citep{munoz18}, filled blue triangles: NGC~6440 \citep{munoz17}, filled orange squares: bulge field stars \citep{vanderswaelmen16}, filled  grey squares: Halo and disk stars \citep{fulbright00,francois07,reddy06,barklem05,venn04}.}
  
  \label{baeu}
  \end{figure}


\subsection{Neutron-Capture elements}
We measured the abundances for five neutron-capture elements: Zr, Ba, Ce, Nd, and Eu. 

As seen in Figure \ref{heavy}, these elements show a gradual decrease with increasing metallicity. This effect is due to the enrichment of iron from SN Ia \citep{vanderswaelmen16}.

The five elements show  good agreement with the bulge field star trend, basically solar abundance for the case of Ba and Eu, and sub-solar for the case of Nd and Zr with a value around -0.3 dex (see Figure \ref{heavy}).

Comparing our results from NGC~6553 with NGC~6528 \citep{munoz18}, we observe a good agreement for Zr and Ba. But, we find a significant difference in  Eu of 0.23 dex. 

The ratio of [Ba/Eu] shown  in Figure \ref{baeu}, is a good indicator of the contribution of the s-process vs. r-process during the evolution of our Galaxy. In this plot we notice an increase of the [Ba/Eu] vs. [Fe/H]  for the bulge field stars, suggesting some contribution from the AGB stars around the solar metallicity \citep{vanderswaelmen16}.

We noticed that the results for our studies with NGC~6440 \citep{munoz17}, NGC~6528 \citep{munoz18} and NGC~6553 (this study) are compatible with the bulge field stars trend. Specifically, regarding NGC~6553, we note a nucleosynthetic history dominated by s-process, indicating an area of formation mainly enriched by AGB stars at the early epoch \citep{vanderswaelmen16}.

\section{Conclusions}
In this article, we have derived detailed chemical abundances for the  GC NGC~6553 from seven RGB star members. Using FLAMES-UVES data, we  measured  the chemical abundances of 20  elements, together with an accurate measurement of the errors. We have performed a detailed comparison with other bulge GCs studied homogeneously as part of our previous studies \citep{munoz17,munoz18} and also compared to results in the literature for other bulge GCs (NGC~6441 and HP~1), as well as for  field stars from the Halo, Disk and the Bulge.

Summarizing the most import results, NGC~6553 is one of the most metal-rich among Galactic GCs; the mean in metallicity found in our sample is  [Fe/H]=-0.14$\pm$0.07 dex, and is homogeneous in iron content. 

Using the alpha-elements Mg, Si, Ca and Ti we obtain the mean of  [$\alpha$/Fe]=$0.11\pm0.05$. Overall, the $\alpha$-elements, iron peak elements and heavy elements measured for NGC~6553, show a good agreement with the bulge field stars trend as we can see in Figures \ref{iron-ele},\ref{alpha1},\ref{alpha2} and \ref{heavy}. Although, it is possible to observe in this GC some compatibility with the disk trend (see Figure \ref{alpha1}),  in agreement with the finding by \citet{zoccali01}. However, we found very good accordance with NGC~6528, another bulge GCs, and with the general chemical patterns of the bulge.

Our most important finding is of a vertical Na-O relation, with a significant intrinsic spread in Na, but almost nonexistent in the case of Oxygen. This is compatible with the other bulge GCs NGC~6528 and NGC~6440 from our previous studies \citep{munoz17,munoz18}. This short extension in the Na-O anticorrelation found in these clusters (NGC~6553, NGC~6528 and NGC~6440) is in agreement with that found by Carretta in his previous studies \citep{carretta15,carretta11,carretta10b} regarding the mass of the GCs. Nonetheless, Carretta mentions other factors that may play a role in this regard. Metallicity is another important factor, considering that these three GCs are metal-rich among galactic GCs, with a metallicity between  [Fe/H]=-0.50 to -0.14 dex, this would be in agreement with what was mentioned by \citet{carretta09b} and \citet{gratton10,gratton11} about the extension of the Na-O and the metallicity. Other factors must come into play, such as the environment  of formation and evolution of these GCs, taking into account that these three metal rich GCs are  members of  the bulge of our galaxy. It is also important to note  that our samples are small, only seven stars in each bulge GCs, therefore we need to increase it to be conclusive about our finding.


Likewise, we have found no Mg-Al anti-correlation, similar to the case of NGC~6528 \citep{tang17}. Finally, we detect the presence of  MPs in this bulge GCs mainly via the spread in Na and Al (see Figure \ref{mg-al-na}).

We measured five neutron capture elements, which follow the trend of the bulge field stars and the bulge GCs from our previous studies (NGC~6440 and  NGC~6528). [Ba/Eu] versus [Fe/H] is dominated by s-process material, indicating a  formation mainly enriched by AGB stars at an early epoch.

Finally, we have presented in this research a new chemical tagging for the  GC NGC~6553. Together with the other two bulge GCs we have studied, these provide a homogeneous dataset for more than a single bulge GC. Clearly the bulge deserves much more dedicated studies to uncover the many  hidden secrets it must contain about the nature of the formation and evolution of this primary Galactic component. Such a study recently begun is the CAPOS (bulge Cluster APOgee Survey) project designed  to observe the bulk of the bulge GCs using the high resolution, near IR multiplexing capabilities of the APOGEE spectrograph.

\section{Acknowledgements}

This work is Based on observations collected at the European Organisation for Astronomical Research in the Southern Hemisphere under ESO programme ID 093.D-0286. SV gratefully acknowledges the support provided by Fondecyt reg. n. 1170518. D.G. gratefully acknowledges support from the Chilean Centro de Excelencia en Astrof\'isica
y Tecnolog\'ias Afines (CATA) BASAL grant AFB-170002.
D.G. also acknowledges financial support from the Direcci\'on de Investigaci\'on y Desarrollo de la Universidad de La Serena through the Programa de Incentivo a la Investigaci\'on de
Acad\'emicos (PIA-DIDULS). We would also like to thank the referee for his valuable comments.

\bibliographystyle{mnras}
\bibliography{biblio.bib}
\bsp	
\label{lastpage}
\end{document}